\documentclass[10pt,conference,twocolumn]{IEEEtran}
\usepackage{filecontents}
\usepackage[noadjust]{cite}
\usepackage{caption}
\usepackage{amsfonts}
\usepackage[dvips]{color, graphicx}
\usepackage[cmex10]{amsmath}
\usepackage{pgfplotstable}
\usepackage{array}
\usepackage{booktabs}
\usepackage{times}
\usepackage[normalem]{ulem} 
\usepackage[lined,boxed,commentsnumbered]{algorithm2e}
\usepackage{flushend}
\ifCLASSOPTIONcompsoc
\usepackage[caption=false,font=normalsize,labelfon
t=sf,textfont=sf]{subfig}
\else
\usepackage[caption=false,font=footnotesize]{subfi
g}
\fi

\DeclareMathOperator*{\E}{E}

\long\def\symbolfootnote[#1]#2{\begingroup%
\def\thefootnote{\fnsymbol{footnote}}\footnote[#1]{#2}\endgroup}

\begin{document}
\title{On the Polar Code Encoding in Fading Channels}
\author{\IEEEauthorblockN{Rui Deng\IEEEauthorrefmark{1},
Liping Li\IEEEauthorrefmark{1},
Yanjun Hu\IEEEauthorrefmark{1}}
\IEEEauthorblockA{\IEEEauthorrefmark{1}
Key Laboratory of Intelligent Computing and Signal Processing of the Ministry\\
of Education of China, Anhui University, China,\\
Email: dengrui@ahu.edu.cn, liping\_li@ahu.edu.cn,  yanjunhu@ahu.edu.cn}}

\maketitle
\begin{abstract}
Besides the determined construction of polar codes in BEC channels, different construction
techniques have been proposed for AWGN channels. The current state-of-the-art algorithm starts with
a design-SNR (or an operating SNR) and then processing is carried out to approximate each individual
bit channel. However, as found in this paper, for fading channels, an operating SNR can not be directly
used in approximating the bit channels.
To achieve a better BER performance, the input SNR for the polar code construction in fadding channels is derived. A selection of the design-SNR for both the AWGN and the fading channels from
an information theoretical point of view is studied.
Also presented in this paper is the study of sacrificing a small data rate to gain orders of  magnitude increase
in the BER performance.
\end{abstract}

\section{Introduction}\label{sec_ref}
Polar codes have been proposed by Erdal Ar$\i$kan in \cite{arikan_iti09}. The fact is that polar codes can achieve the channel capacity at a low encoding and decoding complexity of $\mathcal{O}(N \log N)$. The original format of polar codes in \cite{arikan_iti09} is non-systematic. Later, a systematic version of polar codes was invented by Ar$\i$kan in \cite{arikan_icl11}, which is shown to outperform the non-systematic polar codes in terms of the BER performance.


Channel combining and splitting realize the polarization of $N$ channels. After these two stages, channels are polarized in the sense that bits transmitting in these channels either experience almost noiseless channels or almost completely noisy channels for a large $N$.
Then one can easily achieve a rate of transmission close to the channel capacity, simply by choosing to transmit over the good bit channels and fix the bits in bad channels. However, at a finite block length $N$ and a data rate $R = K/N$ ($K$ being the number of information bits in each codeword of length $N$). A ranking algorithm of the bit channels according to their bit error rate (BER) becomes necessary to select $K$ good channels out of the $N$ channels.
This selection of bit channels completely defines a polar code for any given underlying channel.

In this paper, we consider the underlying channels to be AWGN channels and fading channels.
To select the good indices in a AWGN channel, as stated in \cite{arikan_iti09}, a Monte-Carlo simulation can be
used to sort the bit channels. This algorithm has a high complexity of $\mathcal{O}(MN \log N)$,
where $M$ is the number of iterations of the Monte-Carlo simulation. A more recent construction algorithm for AWGN channels
is proposed by Tal and Vardy \cite{vardy_it13} which is based on an earlier proposal from Mori and Tanaka \cite{mori_icl09} \cite{mori_isit09}. The algorithm in \cite{vardy_it13} is by far the best construction algorithm with  a complexity of $\mathcal{O}(N\cdot\mu^{2}\log\mu)$ where $\mu$ is a user defined variable (Note that this complexity excludes
the quantization of a AWGN channel to a discrete-alphabet channel). In \cite{trifonov_itc12}, the estimation of bit-channels based on a Gaussian approximation is proposed.  This was found to well-approximate the actual bit-channels of polar codes in \cite{li_tsc13} and \cite{wu_icl14}. A similar algorithm after the original proposal in \cite{trifonov_itc12} was studied in \cite{li_tsc13}, \cite{wu_icl14}. The Gaussian approximation
algorithm involves relatively higher complexity function computations, with a complexity of $\mathcal{O}(N)$ function computations (excluding the selection of K best among N metrics obtained).

From the discussions of the constructions of \cite{arikan_iti09, vardy_it13, trifonov_itc12, li_tsc13, wu_icl14}
for AWGN channels, it's seen that all algorithms start with a specified value of the signal-to-noise ratio (SNR).
In practice, it's difficult (or expensive) to compute the code indices with the variation of the SNR. Hence, a design-SNR is desired to construct polar codes that produce a good BER performance for a given data rate $R$ and possible variations
of the system SNR. In \cite{Vangala_15}, a design-SNR is produced from extensive simulations.

In this paper, we propose a selection of the design-SNR for any data rate $R$  from an information theoretical analysis.
For a system with an operating SNR and a data rate $R$, a design-SNR can be used to obtain the code indices
instead of the operating SNR. Both AWGN channels and block fading channels are considered in the analysis of this paper.
In block fading channels, we also conduct a study of sacrificing a small data rate in the large-SNR region
to gain orders of magnitude increase of the BER performance. Simulation results are provided to verify the validness of the selection of the design-SNR and the great increase of the BER performance at a small cost of the data rate.


The rest of the paper is organized as follows. In Section \ref{sec_background}, we describe fundamentals of non-systematic polar code and systematic polar code.
Section \ref{sec_algorithm} introduces a transmission scheme which
can improve the BER performance in orders of magnitude at a small cost to the data rate in fading channels. In Section \ref{sec_choice}, we analyze the design-SNR and present a selection of the design-SNR for both the AWGN and the fading channels. We finally present our simulation results in Section \ref{sec_result} and conclude this paper in Section \ref{sec_con}.

\section{Polar Code Fundamentals}\label{sec_background}
In this section, the relevant theories
about non-systematic polar codes and systematic polar codes are presented.
\subsection{Non-Systematic Polar Codes}\label{sec_non-sys}
A polar code may be specified completely by $(N,K,{\mathcal{A}},u_{{\mathcal{A}}^{c}})$, where $N$ is the length of a codeword, $K$ is the number of information bits encoded per codeword, ${\mathcal{A}}$ is a set of $K$ indices called information bit locations from $\{1,2,....,N-1,N\}$, and $u_{\mathcal{A}^{c}}$ consists of the frozen bits. For an $(N,K,{\mathcal{A}},u_{{\mathcal{A}}^{c}})$ polar code, a codeword is obtained as:
\begin{equation} \label{eq_x1}
\mathbf{x}= {u}_{\mathcal{A}}\cdot G_{\mathcal{A}}+ {u}_{\mathcal{A}^{c}}\cdot G_{\mathcal{A}^{c}}
\end{equation}
where $G_{\mathcal{A}}$ denotes the submatrix of $G$ formed by the rows with indices in ${\mathcal{A}}$, and $G$ equals $F^{\otimes n}$ for any $n=\log_2N$. Here $F$ is the standard polarizing kernel $F=\bigl(\begin{smallmatrix} 1&1 \\ 0&1\end{smallmatrix}\bigr)$.

The Successive Cancellation Decoding (SCD) algorithm in \cite{arikan_iti09} uses a decoding operation that is similar to the belief propagation (BP) algorithm. The likelihoods evolve in the inverse direction from the right to the left, as illustrated with an example in \cite{arikan_iti09}, using a pair of likelihood transformation equations. Then the bit decisions are made at the left of the circuit and is used in the rest of the decisions.


\subsection{Systematic Polar Codes}\label{sec_sys}
The systematic polar code split the codeword into two parts by writing
$\mathbf{x} = (x_{\mathcal{B}}, x_{\mathcal{B}^{c}})$. So (\ref{eq_x1}) can be written as
\begin{equation} \label{eq_xb}
{x_{\mathcal{B}}}= {u}_{\mathcal{A}}\cdot G_{\mathcal{AB}}+
{u}_{\mathcal{A}^{c}}\cdot G_{\mathcal{A}^{c}\mathcal{B}}
\end{equation}
\begin{equation} \label{eq_xbc}
{x_{\mathcal{B}^{c}}}= {u}_{\mathcal{A}}\cdot G_{\mathcal{AB}^{c}}+ {u}_{\mathcal{A}^{c}}\cdot G_{\mathcal{A}^{c}\mathcal{B}^{c}}
\end{equation}
where the $G=F^{\otimes n}$ and $F=\bigl(\begin{smallmatrix} 1&0 \\ 1&1\end{smallmatrix}\bigr)$, the matrix $G_{\mathcal{AB}}$ is a submatrix of the generator matrix
with elements $(G_{i,j})_{i \in \mathcal{A}, j \in \mathcal{B}}$, and similarly for the other submatrices.
The proposition in \cite{arikan_icl11} says that if (and only if) $\mathcal{A}$ and $\mathcal{B}$ have the same number of elements and $G_{\mathcal{AB}}$ is invertible, there exists a systematic encoder $(\mathcal{B},u_{\mathcal{{A}}^{c}})$. It performs
the mapping ${x}_{\mathcal{B}} \mapsto {\mathbf{x}}=({x}_{\mathcal{B}},{x}_{{\mathcal{B}^{c}}})$. The vector $u_{\mathcal{A}}$ can be obtained by computing
\begin{equation} \label{eq_ua}
u_{\mathcal{A}} = (x_{\mathcal{B}}-u_{{\mathcal{A}^{c}}}G_{\mathcal{A}^{c}\mathcal{B}})(G_{\mathcal{AB}})^{-1}
\end{equation}
In \cite{arikan_icl11}, we've known that $\mathcal{B}=\mathcal{A}$ is the necessary conditions of  establishing the one-to-one mapping $x_{\mathcal{B}} \mapsto u_{\mathcal{A}}$. In the rest of the paper, the systematic encoding of poalr codes will use $\mathcal{A}$, instead of $\mathcal{B}$. So, the mapping of (\ref{eq_xb}) and (\ref{eq_xbc}) can be rewritten as:
\begin{equation} \label{eq_xb1}
{x_{\mathcal{A}}}= {u}_{\mathcal{A}}\cdot G_{\mathcal{AA}}+
{u}_{\mathcal{A}^{c}}\cdot G_{\mathcal{A}^{c}\mathcal{A}}
\end{equation}
\begin{equation} \label{eq_xbc1}
{x_{\mathcal{A}^{c}}}= {u}_{\mathcal{A}}\cdot G_{\mathcal{AA}^{c}}+ {u}_{\mathcal{A}^{c}}\cdot G_{\mathcal{A}^{c}\mathcal{A}^{c}}
\end{equation}

\section{Polar Codes Transmission over \\Fading Channels}\label{sec_algorithm}
In this section, we present a transmission scheme which can improve the BER performance in orders of magnitude at
a small cost to the data rate.

The underlying channel we consider is
\begin{equation} \label{eq_yhn}
y = h\cdot {\tilde{x}}+n
\end{equation}
where $h$ follows a normal distribution $\mathcal{N}(0,1)$, $n$ is the additive white Gaussian noise with
mean 0 and variance $\sigma^{2}$, and
$\tilde{x}$ is the binary modulated transmitted symbol: $\{0,1\} \rightarrow  \{1,-1\}$. The
encoded bits $\mathbf{x} = \mathbf{u}G$ is thus modulated to $\mathbf{\tilde{x}}$.
The $N$ channels experienced by $\mathbf{\tilde{x}}$, denoted as $\mathbf{h}$,
are assumed to be i.i.d and are stable for $N_b$
code blocks. This is a block fading channel model which can be observed in various wireless
communication scenarios. In our study, assume $\mathbf{h}$ is known at the receiver.
With the channel estimation of $\mathbf{h}$, the underlying channel
in (\ref{eq_yhn}) can be converted to:
\begin{equation} \label{eq_yc}
\tilde{y} = |h| \cdot \tilde{x} + n
\end{equation}

Due to the limitations of the channel estimation algorithms,
there is a limit of the channel values below which the estimation is unreliable. In this paper, we
do not attempt to discuss the limitations of various channel estimation algorithms. Instead, we
keep this limit of channel values as a variable $\alpha$. As seen in the sequel, the change
of $\alpha$ with channel estimation algorithms does not increase the complexity of our algorithm.

\begin{figure}
{\par\centering
\resizebox*{3.0in}{!}{\includegraphics{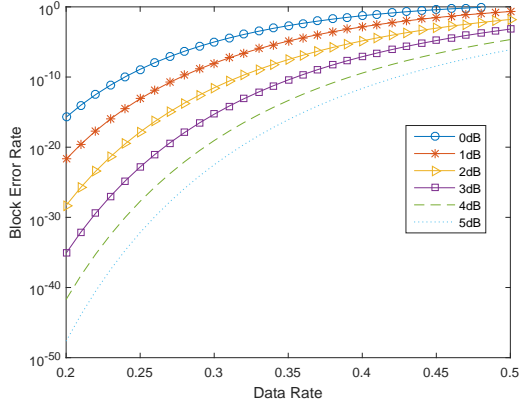}} \par}
\caption{The trend of the block error rate for $N = 1024$ using the asymptotic analysis in (\ref{eq_asymptotic}).}
\label{fig_error_trend}
\end{figure}

For a given $\alpha$, a percentage can be calculated: $p = \Pr\{|h| \le \alpha\}$, which
translates to the fact that $\lfloor N * p\rfloor$ observations at the receiver
are not reliable. These unreliable observations in return will cause a poor
BER performance. In this case, to maintain a given BER performance, either a
higher SNR or a lower data rate can be applied. Here a question arises: whether
the SNR should be increased or the data rate should be lowered? Which selection
yields a better performance? To answer this question, we resort to the work of \cite{hassani_iti13}
for the asymptotic behavior of polar codes. The relationship of the block error rate $P_e$ with
the data rate $R$ and the capacity of the underlying channel $I(W)$ is
rewritten as below from \cite{hassani_iti13}
\begin{equation}\label{eq_asymptotic}
P_e = 2^{-2^{\frac{n}{2}+\sqrt{n}Q^{-1}(\frac{R}{I(W)})+o(\sqrt{n})}}
\end{equation}
where $Q^{-1}(x)$ is the reverse function of $Q(x) = \frac{1}{\sqrt{2\pi}}\int_x^\infty e^{-\frac{t^2}{2}} \, \mathrm{d}t$.
The fundamental information theoretical analysis yields the channel capacity $I(W) = C$ of (\ref{eq_yhn}) as
\begin{equation}\label{eq_ch}
C = \E_{|h|}\{ C(|h|\sqrt{\gamma)} \}
\end{equation}
where $\gamma$ is the SNR and  $C(|h|\sqrt{\gamma})$ is defined as
\begin{equation}\label{eq_c}
\begin{split}
C(|h|\sqrt{\gamma}) &= \int_{-\infty}^\infty\frac{1}{\sqrt{2\pi}}e^{\frac{-(y-|h|\sqrt{\gamma})^{2}}{2}}\log_2 {\frac{2}{1+e^{-2y|h|\sqrt{\gamma}}}}{\rm d}y\\
& +\int_{-\infty}^\infty\frac{1}{\sqrt{2\pi}}e^{\frac{-(y+|h|\sqrt{\gamma})^{2}}{2}}\log_2 {\frac{2}{1+e^{2y|h|\sqrt{\gamma}}}}{\rm d}y
\end{split}
\end{equation}
As the evaluation of (\ref{eq_c}) is not closed, we can use an upper bound of $C \le C(\E\{|h|\}\sqrt{\gamma})$ to
approximate $I(W)$. The function $P_e$ is concave in terms of the data rate $R$ when fixing $I(W)$.
It's also concave in terms of $SNR$=$\gamma$ for a fixed data rate $R$.
Fig.~\ref{fig_error_trend} shows $P_e$ as a function of the data rate $R$ under several SNR values.
From Fig.~\ref{fig_error_trend}, we can see clearly that with a fixed data rate, increasing SNR
does not yield a significant BER performance increase at high SNR regions. Instead, decreasing
the data rate $R$ is a better choice. With this principle, we propose the following Algorithm \ref{algorithm_a}
to improve the BER performance in fading channels whenever a certain amount
of bad channels are detected. In this algorithm, $Q_s$ is a channel index matrix chosen according to the quality of bit channels and $M$ is approximately equal to $N \cdot p$.

\IncMargin{1em}
\LinesNumbered
\begin{algorithm}
\SetKwInOut{Input}{INPUT} \SetKwInOut{Output}{OUTPUT}
\Input{a matrix $h = \{h_1,h_2,\cdots,h_N\}$, $\alpha$, data rate $R$, $N$, $SNR$, $M$, $Q_s$}
\Output{$chosen\_coding\_idx$ and $chosen\_frozen\_idx$ \tcp*[h]{new good channel indices and new bad channel indices}}
\BlankLine
$n\_info = N * R$ \;
Take the absolute value of $h(coding\_idx)$ to $h\_abs$ \;
$rm\_Num = 0$ \;
\For {$i \leftarrow 1$ \KwTo $n\_info$}{
        \If(\tcp*[h]{accumulate the number of $h$ samller than $\alpha$}) {$h\_abs(i) < \alpha$}{
               $rm\_Num = rm\_Num+1$ \;
            }
       }
\If {$rm\_Num <= M$}{
            $Mi = rm\_Num$ \;
}
\If {$rm\_Num > M$}{
            $Mi = M$ \;
}
$c$ $\leftarrow$ channel capacity of the $SNR$ \;
Round down $Mi * c$ to $Num$\;
$chosen\_coding\_idx = Q_s(1, (n\_info-Num))$ \;
$chosen\_frozen\_idx = Q_s(n\_info-Num+1, N)$ \;
$chosen\_coding\_idx$ and $chosen\_frozen\_idx$ ranked by the order from small to large \;

\caption{ The algorithm
to obtain new good channel indices and bad channel indices in fading channels whenever a certain amount
of bad channels are detected}
\label{algorithm_a}
\end{algorithm} \DecMargin{1em}

With Algorithm \ref{algorithm_a}, a system in a fading channel first selects a $\alpha$ according
to the channel estimation algorithm. Then another value  $M = \lfloor Np \rfloor$ is selected which can be
a tradeoff between the BER performance and the data rate decrease. When $M$ (or any number smaller than
$M$) unreliable channel estimations are detected, a data rate decrease is performed: $M*c$ channels are
changed from the data-bearing channels to the frozen channels. Here $c$ is the calculated channel
capacity according to the operating SNR $=\gamma$ and the capacity of $C=C(\E\{|h|\}\gamma)$ in (\ref{eq_c}).
This decrease of $M*c$ is that the $M$ unreliable observations theoretically lose $M*c$ information
bits.

It's worthwhile to emphasize that the  variation of $\alpha$ and $M$ does
not increase the complexity of Algorithm \ref{algorithm_a}. This resides in the choice of the sorted
channels $Q_s$. A change of  $\alpha$ and $M$ only changes the number $M*c$ which eventually only results
in how many entries are selected from $Q_s$. As long as $Q_s$ is provided, the complexity does not
change with the variation of the parameter $\alpha$ and $M$. In the next Section, we provide
a theoretical way to construct $Q_s$ for a given data rate $R$.


%
%


\section{Design-SNR Analysis}\label{sec_choice}

For a system with an operating SNR = $\gamma$ and a data rate $R$ in AWGN channels,
a construction based on \cite{vardy_it13} can be carried out with this $\gamma$.
Theoretically, with any change in SNR, a new construction should be obtained to
accommodate this change of SNR. However, to a lot of
systems, a real time construction is too complicated to perform.
In \cite{Vangala_15}, a design-SNR is found from simulations which can produce a good BER performance
for a range of SNRs. In this paper, we provide an information theoretical foundation for the
existence of the design-SNR for a fixed data rate $R$.

Let's first look at the AWGN channels. The constellation constrained channel capacity of the AWGN
channel is shown in (\ref{eq_ch}) with $h=1$. For a fixed data rate $R$, the required SNR $=\gamma$
to achieve this data rate $R$ can be obtained as:
\begin{equation}\label{eq_gamma_1}
\gamma = C^{-1}(R)
\end{equation}
where $C^{-1}(x)$ is the inverse function of $C(x)$ in (\ref{eq_c}). Let's denote the required
SNR for a data rate $R$ as $\gamma_R$. For a fixed data rate $R$, this $\gamma_R$ is the
design-SNR which should be used for the construction of polar codes.

For fading channels, in order to use the construction algorithm like \cite{vardy_it13}, we need
to convert the channel in (\ref{eq_yc}) to an AWGN channel. One way to do this is to take
the mean of the channel in (\ref{eq_yc}) with respect to $h$
\begin{equation}
\E_{h}\{\tilde{y}\} \approx \E\{|h|\} \cdot \tilde{x} + n
\end{equation}
which has an equivalent $SNR$ of
\begin{equation} \label{eq_x2}
\gamma_h = E\{|h|\}^{2}\gamma =\mu^{2}\gamma.
\end{equation}
In (\ref{eq_x2}), $\mu = \E\{|h|\}$. For the normal distribution $h$, $\mu \approx 0.4$. Therefore,
we have
\begin{equation} \label{eq_gamma}
\gamma_h ~(\text{dB}) = \gamma ~(\text{dB}) - 8 ~(\text{dB})
\end{equation}
If a point-by-point construction is to be carried out for the underlying channel in (\ref{eq_yhn}), then
the input SNR for the construction algorithm should be a modified version $\gamma_h$ shown in (\ref{eq_gamma}) instead
of directly using the operating SNR $\gamma$. This point is verified in the simulation results in the next section: the BER
performance of polar codes constructed using $\gamma$ is poorer than that constructed by $\gamma_h$.
%
%

Like AWGN channels, a design-SNR also exists for fading channels. However, from
the simulation results, we find that the design-SNR does not follow the relationship
shown in (\ref{eq_gamma}): a design-SNR of $\gamma_R$ for a AWGN channel does not
produce a design-SNR $\gamma_{Rh} = \gamma_R - 8$ (all in dB) for the fading channel in (\ref{eq_yhn}).
Surprisingly, AWGN channels and the fading channels  in (\ref{eq_yhn}) share the same design-SNR.
Simulation results are provided to show the performance of the design-SNR selections for AWGN
channels and fading channels.



\section{Numerical Results}\label{sec_result}
To verify the fading channel transmission scheme of polar codes in Algorithm \ref{algorithm_a} and the design-SNR selection,
simulation results are provided in this section. All the polar codes construction in this section are based on
\cite{vardy_it13}.

\begin{figure}
{\par\centering
\resizebox*{3.0in}{!}{\includegraphics{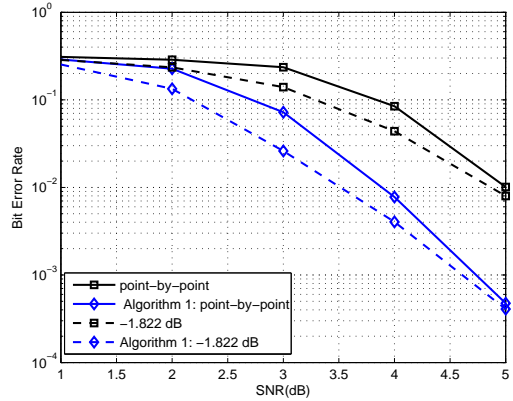}} \par}
\caption{The BER performance of systematic polar codes transmission over fading channels using Algorithm \ref{algorithm_a}.
The block length is $N = 1024$ and the data rate $R = 0.36$.}
\label{fig_algorithm}
\end{figure}

Algorithm \ref{algorithm_a} is carried out using the following parameters: $N=1024$, $R=0.36$, $\alpha = 0.2$, and $M=64$.
The sorted channel indices $Q_s$ are generated in two ways: 1) point-by-point construction of (\ref{eq_gamma});
2) the design-SNR for the data rate $R=0.36$. The design-SNR is computed using (\ref{eq_gamma_1}). For $R=0.36$,
the desgin-SNR is computed to be $-1.822$ dB. The BER performance of systematic polar codes are shown in Fig.~\ref{fig_algorithm}, where the solid squared line is the original BER performance with code
indices selected from the point-by-point construction with the input SNR given in (\ref{eq_gamma}) and the solid
line with diamonds is the BER performance of polar codes constructed this way with Algorithm \ref{algorithm_a} running
at the same time.
Clearly, the BER performance is greatly improved with Algorithm \ref{algorithm_a}: the BER is reduced by 21 times
while the data rate is only reduced by 15\%. The two dashed lines in Fig.~\ref{fig_algorithm} are
the corresponding BER performance using the design-SNR -1.822 dB instead of a point-by-point construction. The same
improvement is observed when running our Algorithm \ref{algorithm_a}.

\begin{figure}
{\par\centering
\resizebox*{3.0in}{!}{\includegraphics{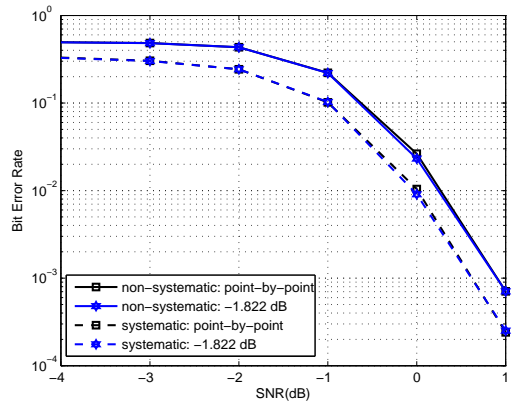}} \par}
\caption{The design-SNR performance in AWGN channels. The polar code have a block length $N=1024$ and
the data rate $R=0.36$.
}
\label{fig_awgn_comparison}
\end{figure}

Fig.~\ref{fig_awgn_comparison} shows the BER performance comparison between the construction using the design-SNR and
the point-by-point construction in AWGN channels. The polar code has a block length $N=1024$ and $R=0.36$. The design-SNR for this
case is the same as in Fig.~\ref{fig_algorithm}.
The two solid lines in Fig.~\ref{fig_awgn_comparison} are the non-systematic
BER performance using the point-by-point construction and the design-SNR construction.
From Fig.~\ref{fig_awgn_comparison}, it can be seen that the BER performance of non-systematic polar codes
using the design-SNR construction matches closely with the point-by-point construction. The same can be seen
for the systematic polar codes.

\begin{figure}
{\par\centering
\resizebox*{3.0in}{!}{\includegraphics{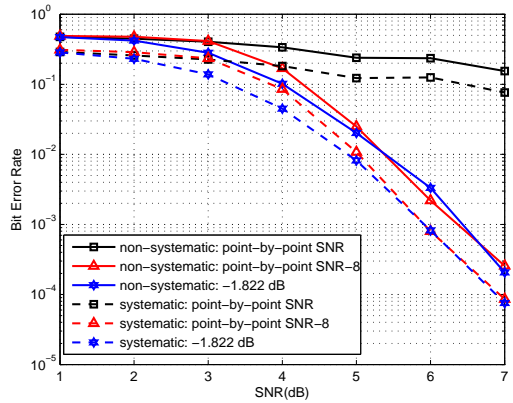}} \par}
\caption{The design-SNR performance of polar codes in fading channels. The code
block length is $N=1024$ and the data rate $R=0.36$.}
\label{fig_simulation_result1}
\end{figure}

The simulation results of the design-SNR in fading channels are shown in Fig.~\ref{fig_simulation_result1}.
The polar code block length is again $N=1024$ and the data rate is $R=0.36$.
According to Section \ref{sec_choice}, in fading channels, the design-SNR is the same as that of the AWGN
channels. Therefore in this case, the design-SNR is -1.822 dB. The two square lines (solid and dashed) are
the BER performance of non-systematic and systematic polar codes constructed point-by-point using the operating
SNR. The two lines (solid and dashed) with up-triangles  are the BER performance of non-systematic and
systematic polar codes constructed point-by-point using the operating $\text{SNR - 8}$ as shown in (\ref{eq_gamma}).
And the two lines (solid and dashed) with hexagrams are the BER performance of non-systematic and
systematic polar codes constructed using the design-SNR $-1.822$ dB.
The first thing to notice from Fig.~\ref{fig_simulation_result1} is that the BER performance of
both the systematic and non-systematic polar codes with the point-by-point construction
using the operating SNR is much worse than the construction using $\text{SNR - 8}$ and the design-SNR of -1.822 dB.
This shows that the point-by-point construction of polar codes in fading channels of (\ref{eq_yhn}) should
use an input SNR in (\ref{eq_gamma}) instead of directly using the operating SNR. In the mean time, the
BER performance of both the systematic and non-systematic polar codes with the design-SNR is either better
or equal to that of the point-by-point construction using SNR - 8 dB. The design-SNR clearly can be
used for the polar code construction in fading channels.

%
%
%

%

\section{Conclusion}\label{sec_con}
In this paper, we propose a fading channel transmission scheme of polar codes which can greatly
improve the BER performance at a small cost to the data rate. An algorithm is given to implement the
proposed scheme. In the mean time, a design-SNR selection is proposed based on an information theoretical
analysis. This design-SNR selection is valid for both the AWGN channels and fading channels. Simulation
results are provided which verified the transmission scheme and the design-SNR selection.

\bibliography{../ref_polar}
\bibliographystyle{IEEEtran}
\end{document}